# Probing superconducting gap in CeH$_9$ under pressure

Zi-Yu Cao,[1,*] Seokmin Choi,[1,*] Liu-Cheng Chen,[2,3] Philip Dalladay-Simpson,[3] Harim Jang,[1] Federico Aiace Gorelli,[3,4,5] Jia-Feng Yan,[6] Soon-Gil Jung,[7] Ge Huang,[3] Lan Yu,[3] Yongjae Lee,[8] Jaeyong Kim,[6] Tuson Park,[1,†] and Xiao-Jia Chen[2,3,9,†]

[1] *Center for Quantum Materials and Superconductivity (CQMS) and Department of Physics, Sungkyunkwan University, Suwon 16419, Republic of Korea*

[2] *School of Science, Harbin Institute of Technology, Shenzhen 518055, China*

[3] *Center for High Pressure Science Technology Advanced Research (HPSTAR), 1690 Cailun Road, Shanghai 201203, P. R. China*

[4] *Shanghai Advanced Research in Physical Sciences (SHARPS), Pudong, Shanghai 201203, P. R. China*

[5] *Istituto Nazionale di Ottica, Consiglio Nazionale delle Ricerche (CNR-INO), Florence 50125, Italy*

[6] *Department of Physics, HYU-HPSTAR-CIS High Pressure Research Center, Hanyang University, Seoul 04763, Republic of Korea*

[7] *Department of Physics Education, Sunchon National University, Suncheon 57922, Republic of Korea*

[8] *Department of Earth System Sciences, Yonsei University, Seoul 03722, Republic of Korea*

[9] *Department of Physics and Texas Center for Superconductivity (TcSUH), University of Houston, Houston, TX 77204, USA*

[*] These authors contributed equally to this work.

[†] Correspondence should be addressed to TP (tp8701@skku.edu) and XJC (xjchen2@gmail.com)

**The recent discovery of superconductivity in hydrogen-rich compounds has garnered significant experimental and theoretical interest because of the record-setting critical temperatures. As the direct observation of the superconducting (SC) gap in these superhydrides is rare, the underlying mechanism behind its occurrence has yet to be settled down. Here, we report a successful synthesis of the *P6₃/mmc* phase of CeH$_9$ that exhibits the SC transition with SC critical temperature of about 100 K at a pressure of about 100 GPa. The observation of the zero electrical resistance and the critical current demonstrates that the SC phase is realized in Ce-based superhydride. Quasiparticle scattering spectroscopy (QSS) reveals the Andreev reflection at zero bias voltage, a hallmark of superconductivity, in the differential conductance. The obtained SC gap-to-$T_c$ ratio of 4.36 and temperature dependence of the SC gap are consistent with the prediction from the Bardeen–Cooper–Schrieffer theory with a moderate coupling strength. The successful realization of QSS under Megabar conditions is expected to provide a desired route to the study of the mechanism of superconductivity as well as the establishment of the SC phase in superhydride high-$T_c$ systems.**

Superconductivity in the hydrogen-based materials has commanded a significant interest owing to the technologically disruptive power of a room-temperature superconductor (RTSC). Metallic hydrogen has long been proposed to realize this elusive state because of its low mass and density-induced quantum effects, resulting in inherently large vibrational frequencies that enhance the superconducting critical temperature ($T_c$) based on the Bardeen–Cooper–Schrieffer (BCS) mechanism [1, 2]. However, the experimental realization of the metallic form of solid hydrogen, a phenomenon proposed in the decade following the discovery of superconductivity, has remained as elusive as RTSC owing to the challenging pressures required (~500 GPa) [3]. As an alternative approach, motivated by chemical pre-compression and the idea of a "doped" metallic hydrogen scenario, incorporation of hydrogen into other materials was proposed to invoke the high $T_c$s of metallic hydrogen in a more experimentally accessible regime [4, 5]. This approach has found recent success, which was first identified in $H_3S$ via theoretical predictions [6-8] with subsequent experimental confirmation of superconductivity with a $T_c$ of 203 K [9], thus promoting hydrogen-bearing systems to the forefront in the efforts to realize RTSC, the holy grail of materials science [10-15].

The isotopic effect and similar $T_c$ values between experimental and theoretical works indicate that the mechanism of superconductivity in hydrogen-bearing materials can be attributed to an electron-phonon coupling interaction [11-13, 16, 17]. This is well demonstrated in a class of materials known as superhydrides, in which a lanthanide series element is encapsulated by a hydrogen cage and viewed as a clathrate-type structure that is a favorable configuration for high $T_c$ values [10, 11]. For example, the $T_c$ of $LaH_{10}$, which is close to room temperature, differs by only ~10 % between the numerical simulations and experimental observations [8, 10, 11]. The creation of superhydrides involves pressures exceeding one million atmospheres and temperatures higher than 1000 K in a diamond anvil cell (DAC). In addition, the limited sample size of ~30 μm trapped within a diamond vice makes it difficult to probe SC properties in the superhydrides. The measurements of zero resistance under field, diamagnetic susceptibility, and Hall coefficient have been crucial in establishing SC phases in a few superhydride superconductors [18-21]. However, the direct investigation of the SC gap which is closely related to the pairing mechanism has been rare, with only an infrared reflectance spectroscopy being conducted on $H_3S$. [22].

Here, we report an experimental study on the quasiparticle scattering spectroscopy (QSS) of $CeH_9$ under megabar environments. The experiments were conducted in two independent cell marked as Cell-1 and Cell-2. The zero electrical resistance and the critical current below $T_c$ show that the SC state is realized in Ce-based superhydride $CeH_9$. Corroborating this conclusion, Andreev reflection, the hallmark of a SC state, was evidenced below $T_c$, where the differential conductance is best described by the Blonder-Tinkham-Klapwijk (BTK) model with an s-wave energy gap of 21.03 meV. The temperature dependence of the SC gap obtained from the QSS agrees well with the *s*-wave gap predicted by the BCS theory. These results provide significant physical insights into characterizing the SC state of this class of high-$T_c$ superhydride materials, thus constraining viable theoretical models for the mechanism of superconductivity.

The SC $P6_3/mmc$ phase of CeH$_9$ has been successfully synthesized at approximately 100 GPa, with each Ce atom bonded to an H$_{29}$ clathrate-like cage structure (see inset of Fig 1c). Figures 1a and b show photographs of Ce, ammonia borane, and the four platinum leads before and after laser heating in Cell-1. Ammonia borane has shown excellent performance in previous research [10] and has been employed as a hydrogen source and a pressure-transmitting medium. After the heat treatment, the pressure was decreased from 101.9 to 93.5 GPa. Clear Raman modes of hydrogen located near 650 cm$^{-1}$ (S1) and 4194 cm$^{-1}$ (vibron) were observed throughout the sample chamber (Fig 1d and Fig S10). Meanwhile, Ce lost its metallic luster and expanded from 25×25 to 30×30 μm$^2$ owing to the absorption of hydrogen. Following a $T_c$ of 98 K, the electrical resistance sharply drops to zero after laser heating (Fig 2a). Raman spectroscopy and X-ray diffraction (XRD) measurements were performed after the realization of superconductivity. Further, 5×5 mappings (5-μm step) of the XRD patterns revealed that the laser-heated sample yielded a mixture of two phases (Fig. S1a). A representative XRD pattern (point 14 in Fig S1a) is shown in Fig 1c, where CeH$_9$ with the space group of $P6_3/mmc$ and CeH$_4$ ($I4/mmm$) were predominant. The lattice parameters obtained by Le Bail fit were $a$ = 2.838(3) Å and $c$ = 5.801(1) Å for the CeH$_4$ phase and $a$ = 3.674(4) Å and $c$ = 5.427(4) Å for the CeH$_9$ phase. The unit-cell volumes per formula unit were 23.37(4) and 31.78(1) Å$^3$ for CeH$_4$ and CeH$_9$, respectively, which are in good agreement with the theoretical calculations and experiments [15, 23, 24].

In the cell-1, a weak peak marked by an asterisk at 164 cm$^{-1}$ was observed in the Raman spectra (Fig. 1d), which could be attributed to a phonon mode of the synthesized superhydride [25]. During the measurements of the SC critical current ($I_c$) and QSS after three months, the onset of $T_c$ showed a increase to 105 K with a considerably broader transition temperature at similar pressures (Fig 2a). In both runs, a two-step SC transition was observed, which was associated with the two SC phases of CeH$_9$ ($P6_3/mmc$) and CeH$_{10}$ ($Fm\bar{3}m$) in Ref. [15]. In our case, however, the CeH$_{10}$ ($Fm\bar{3}m$) phase was either too weak or absent from the XRD results, indicating that this two-step behavior may originate from other source, such as the local pressure gradient. The pressure dependences of $T_c$ of CeH$_x$ superhydrides were summarized in Fig. 2**b**.

Figure 3**a** shows the current-voltage (*I-V*) characteristics in cell-1 within the range of ± 80 mA, which is, to our knowledge, the highest applied current in the superhydride system that allows the identification of the temperature dependence of $I_c$ accurately as it adopts a SC state [12]. The threshold value of current ($I_c$ = 74 mA) at which a measurable voltage first appears was evident in the SC state at 60 K, which decreased with an increase in temperature and was almost zero as the temperature was close to the $T_c$ offset of 78 K. The non-Ohmic behavior in the *I-V* curve, which appears between the $T_c$ offset of 78 K and $T_c$ onset of 120 K, suggests that superconductivity is suppressed in a portion of the sample (Fig. 3**b**). Above $T_c$, the *I-V* characteristics became linear and followed an Ohmic behavior. The absolute value of the differential resistance |d*V*/d*I*| is derived from the *I-V* curves and is plotted as a function of the current *I* in Fig 3**c**. The zero-resistance plateau at finite current in the low-temperature regime demonstrates the expected behavior for superconductivity. The critical current density $J_c$ was estimated by using the approximate dimension of the sample (30×30×1 μm$^3$). As indicated by the solid line in Fig. **3d**, the temperature dependence of $J_c$ was reasonably explained by the $\delta l$-pinning model $J_c(T) = J_c(0) \times (1 - (T/T_c)^2)^{5/2}(1 + (T/T_c)^2)^{-1/2}$ with a $J_c(0)$ of 4 MA/cm$^2$, which could be attributed to spatial fluctuations of the charge carrier mean free path (*l*) [26-29].

Figure 4 illustrates the dependence on the bias voltage of the differential conductance, d*I*/d*V*, at 105 GPa in Cell-2, while the data in Cell-1 is presented in the supplementary information. In contrast to critical current measurements, this approach allows us to study the scattering rate of energized electrons at the junction interface. The relatively high bias voltage, which was applied to cover the large SC gap in the CeH$_9$ superconductor, incurred the complex background originated from various sources [30, 31]. The differential conductance at 2 K is displayed in the inset to Fig. 4a, where the red solid line describes a polynomial function to account for the smooth varying background (see Fig. S7 for details in the SI). Figure 4a representatively shows the normalized differential conductance divided by the background at 2 K, (d$I$/d$V$) / (d$I$/d$V$)$_{BG}$, where the solid line is the best fitting result of the BTK model for an s-wave SC gap of 21.04 meV. Figure 4b selectively shows the temperature dependence of the normalized d$I$/d$V$ from 2 K to 90 K with an offset for clarity. As the temperature increases, the zero-bias conductance peak from the Andreev reflection, the hallmark of a SC state [32, 33], is gradually suppressed and disappears at temperatures above $T_c$ onset of 109 K (Fig. S8).

The temperature evolution of the SC energy gap in CeH$_9$ is summarized in the inset of Fig. 4c, where the gap values were obtained from the BTK analysis of the zero-bias conductance peak for Cell-2. We note that the SC gap value for Cell-1 was qualitatively determined as the deviation point from the background signal (Section C in SI). When the reduced SC gap of Δ(T)/Δ(0) is plotted against the reduced temperature ($T/T_c$), the gap values determined from two independent measurements were collapsed on top of each other, as shown in Fig. 4c. The temperature dependence of the SC gap is reasonably explained by the gray line predicted from the phonon-mediating BCS theory and the gap-to-$T_c$ ratio, 2Δ(0)/$k_B T_c$, is 4.36, suggesting that CeH$_9$ superhydride belongs to a moderately coupled s-wave superconductor. The superconductivity and gap-to-$T_c$ ratio match well with the theory [16, 34], offering direct confirmation of earlier predictions.

In summary, the superconductivity of CeH$_9$ was confirmed through zero electrical resistance and the suppression of the SC transition by the applied critical current. The QSS measurements revealed Andreev reflection spectroscopy near the zero-bias voltage in the differential conductance, which was reasonably explained by the BTK model with an s-wave order parameter. The temperature dependence of the SC gap obtained from the QSS is in good agreement with the *s*-wave gap predicted by the BCS theory. Taken together with the gap-to-$T_c$ ratio of 4.36, these results suggest that the CeH$_9$ superhydride belongs to an s-wave superconductor with a moderate electron-phonon coupling strength. The realization of the QSS technique under the megabar pressure range, which lead to the direct observation of the SC gap in Ce-based superhydride, offers a promising avenue for investigating the mechanism of superconductivity as well as establishing the SC state itself in the superhydride high-$T_c$ materials.

**Figure Captions**

**Fig. 1: Sample synthesis and structure characterization of Ce polyhydrides. a, b** The optical micrograph of the electrode system from different sides of the cell before and after laser heating at the pressure of around 100 GPa. After laser heating, the pressure decreases from 101.9 to 93.5 GPa. **c** Synchrotron XRD patterns of the laser-heated sample at point 14 (Fig. S1) and the Le Bail refinements of $P6_3/mmc$-CeH$_9$ and $I4/mmm$-CeH$_4$. The black open circles are experimental data points, while the red and blue line shows the fittings of CeH$_9$ and CeH$_4$, respectively. The gray line describes the residues. The inset shows the crystal structure and H cage of CeH$_9$ - $P6_3/mmc$, where large green and small pink spheres represent Ce and H atoms, respectively. **d** Raman spectra for the sample before and after laser heating at point 19 (Fig. S1). The insert shows the scaled spectra for clarity.

**Fig. 2: Superconducting critical temperature as a function of pressure for CeH$_x$ superhydrides. a** Electrical resistance as a function of temperature performed in the time interval of three months in Cell-1. The black curve is measured at the time near the measurement of XRD and Raman scattering. The red curve is obtained during the measurement of $I_c$ and QSS. The blue dashed lines represent linear fittings of the normal and transition states, with $T_c$ defined at their intersection. Inset magnifies the resistance near $T_c$, where the zero resistance state is approached at 93 K. **b** Experimental $T_c$ as a function of pressure for CeH$_9$ and CeH$_{10}$. The experimental data in Cell-1 are shown as red stars. The error in pressure is based on the pressure shift in the cooling process. The open symbols are obtained from Ref. 15 for comparison. The $T_c$ from Cell-2 is marked with a purple star.

**Fig. 3: Current-voltage curves at zero field for CeH$_9$. a** $I$-$V$ curves of CeH$_9$ at temperatures below 80 K and pressure of 94.8 GPa. Inset shows the $I$-$V$ curves at temperatures between 80 K and 130 K. Arrows indicate the temperature variation. **b** $I$-$V$ curves at temperatures between 80 K and 130 K. The dashed lines represent the liner Ohmic *I-V* behavior at each temperature. **c** The absolute value of the differential resistance curves, |d*V*/d*I*|, is plotted at several temperatures ranging from 61.5 to 72.2 K. The |d*V*/d*I*| is obtained from the *I-V* relationship of **a** by taking the derivative of *V* with respect to *I*. **d** The temperature dependence of $J_c$ obtained from the *I-V* curves between 60 and 80 K. The $J_c$ was calculated assuming the possible cross-section of the sample of 30×1 μm$^2$. The $\delta l$-pinning model represented by the solid green line describes the experimental data well. The inset shows the fitting of the $\delta l$-pinning model down to 0 K.

**Fig. 4: Quasiparticle scattering spectra under megabar pressure.** **a** Normalized differential conductance, (d*I*/d*V*) / (d*I*/d*V*)$_{BG}$, at 2.0 K as a function of the bias voltage. The solid red line is the best fit for the BTK model with an s-wave order parameter. The inset displays the raw data and the polynomial background at 2.0 K. **b** Temperature evolution of (d*I*/d*V*) / (d*I*/d*V*)$_{BG}$ from 2.0 to 90 K. The dotted lines are the best fits by the modified BTK model. As the temperature approaches $T_c$, the BTK analysis becomes extremely difficult owing to the weak Andreev reflection, as shown in Fig. S8. **c** Temperature dependence of the reduced SC energy gaps obtained from contacts in Cell-1 (red symbols) and Cell-2 (purple symbols) against the reduced temperature at the pressure of 95 and 105 GPa, respectively. The grey line is the prediction from the BCS theory with an s-wave order parameter. The inset shows the SC energy gap obtained from the BTK analysis for Cell-2.

**Figures**

**Figure 1**

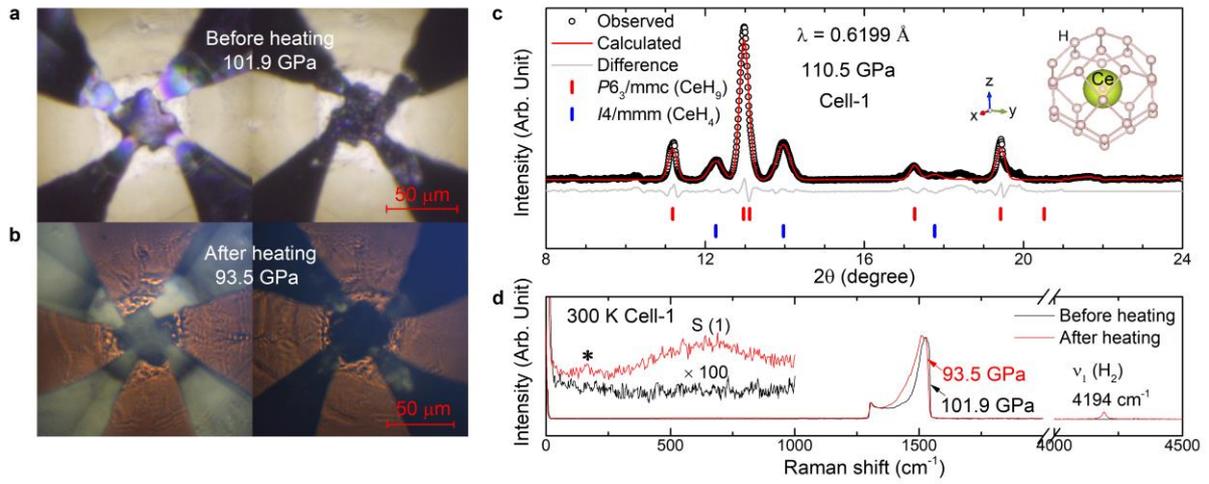

**Figure 2**

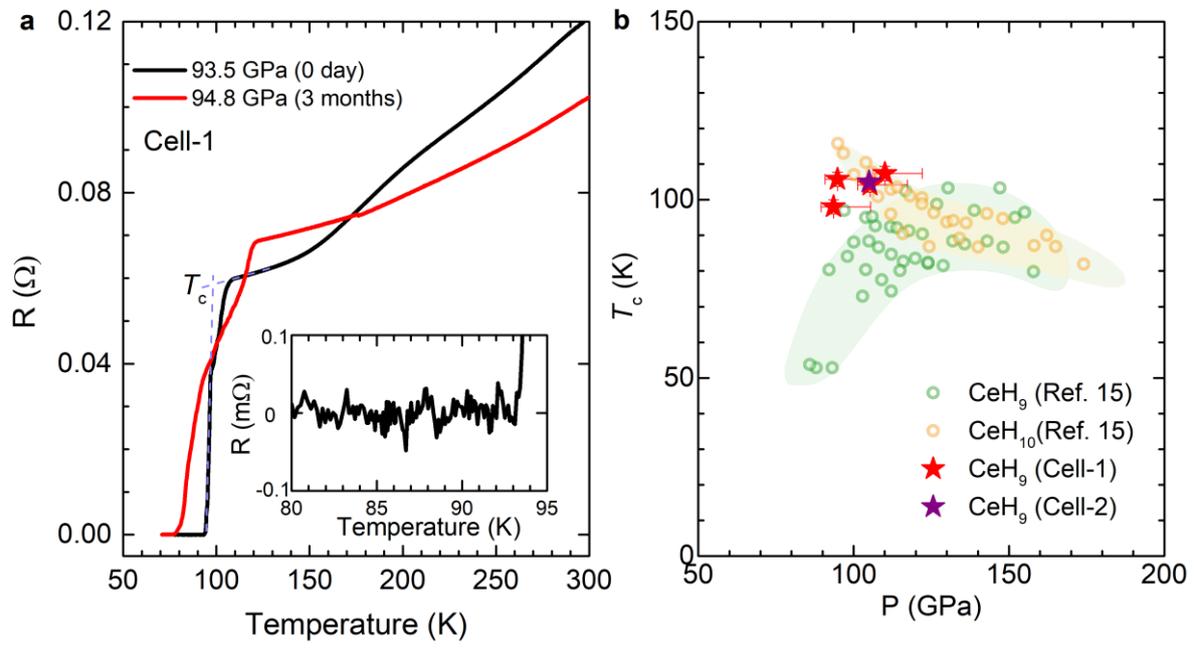

**Figure 3**

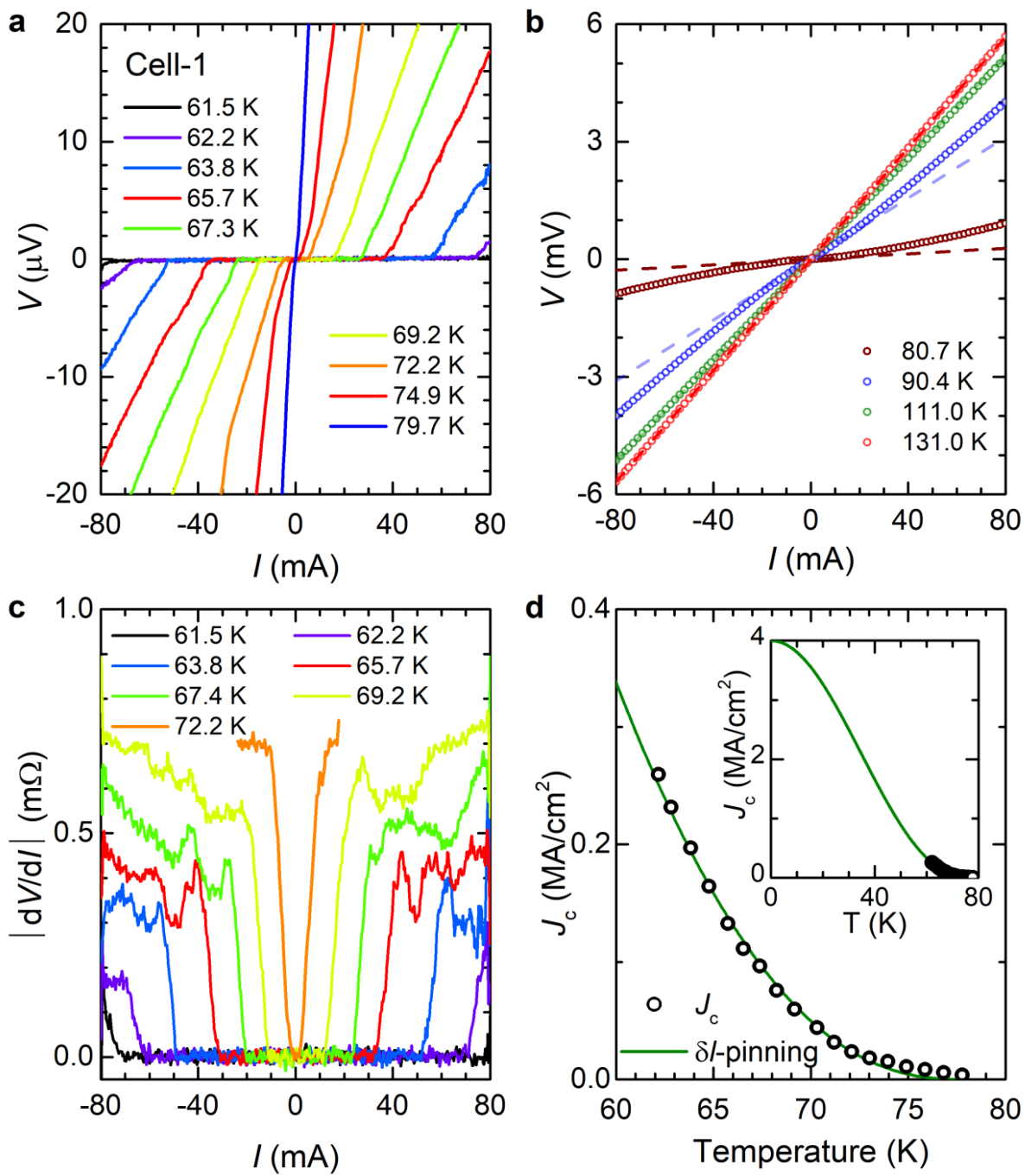

**Figure 4**

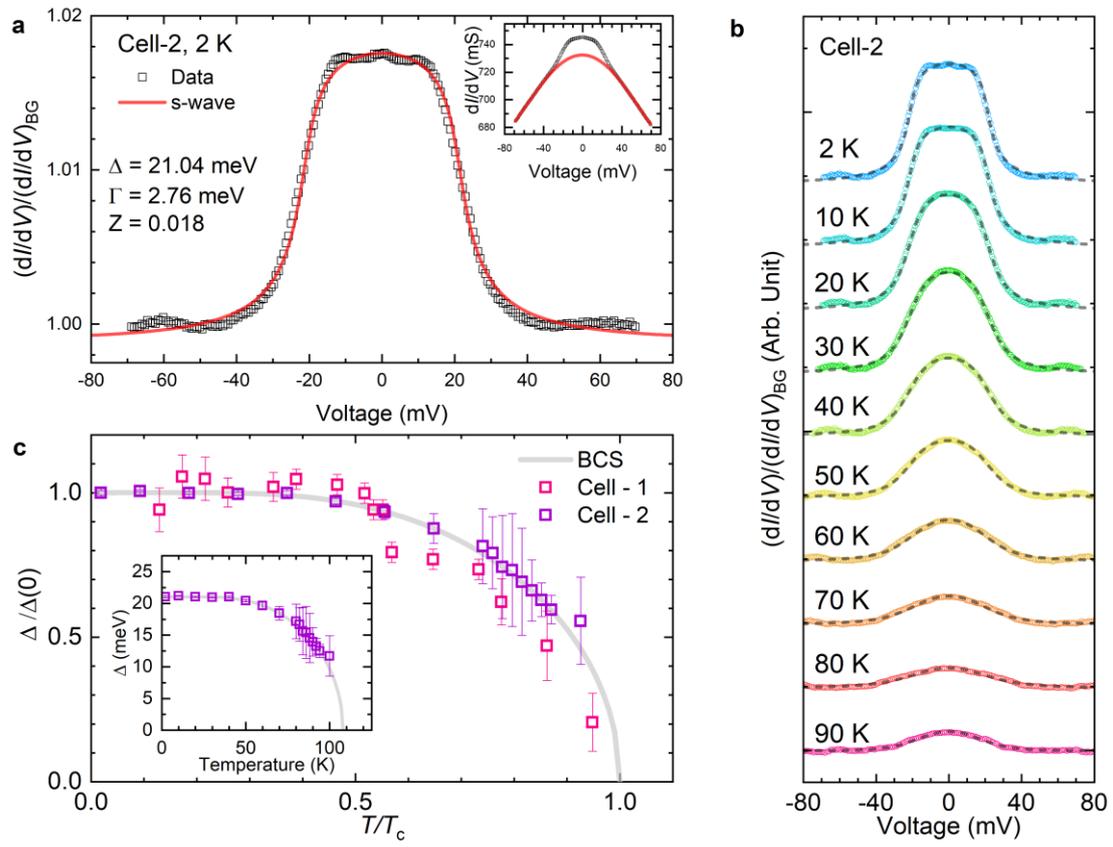

## METHODS

**Diamond anvil cell preparation.**

Two pairs of ultra-low fluorescence Boehler cut diamonds was used in the experiment in Cell-1 and Cell-2. The diamonds had a culet of 100 μm and were beveled at 8.2º to a diameter of 300 μm. The 0.25 mm thick rhenium gasket was electrically insulated from the electrodes by a mixture of epoxy and c-BN. The ammonia borane was employed as both a hydrogen source and a pressure-transmitting medium. Four platinum slices have adhered to a 25×25×1 μm$^3$ Ce (purchased from Alfa Aesar, purity 99.8%) with van der Pauw configuration [35]. The tip of the Pt was flattened to less than 1 μm in advance to avoid further broadening under pressure. The total resistance between electrodes was typically 5-6 Ohm before laser heating and 4-5 Ohm after laser heating. Both the ammonia borane and Ce were loaded in the argon-protected glovebox. The pressure was determined by the position of the high-frequency diamond Raman signal [36]. A symmetry pressure cell and a 23 mm diameter BeCu cell were employed in Cell-1 and Cell-2, respectively.

**Sample synthesis.**

The CeH$_9$ was synthesized by laser heating under megabar pressure. The SPI 100W air-cooled fiber laser with a wavelength of 1064 nm to a spot size of approximately 5 μm in diameter was used in laser heating. In the synthesis, H$_2$ was provided by the ammonia borane. When completely dehydrogenated, one mole of ammonia borane yields three moles of H$_2$ plus insulating c-BN [10]. Both sides of the sample were heated, and the focal point of the laser was raster scanned across the sample to ensure a thorough synthesis of the hydride material. A clear, glowing spot can be observed during laser heating. After laser heating, a Raman vibron of hydrogen located near 4200 cm$^{-1}$ was observed all over the sample and Ce lost the metallic luster and was expended from 25×25 to 30×30 μm$^2$ due to hydrogen absorption.

**Structural characterization.**

In situ high-pressure synchrotron X-ray diffraction patterns from the 5×5 mapping with a 5 μm step, covering the region where Raman spectra were acquired, were recorded at the 15U1 beamline of Shanghai Synchrotron Radiation Facility with the wavelength of 0.6199 Å and a beam spot of ~5-8 μm. The appearance of the spotty powder diffraction rings indicates melting and recrystallization, followed by the formation of relatively large crystallites of a new phase. Two-dimensional X-ray diffraction patterns were integrated and analyzed using the Dioptas software package [37]. The analysis of the diffraction patterns and calculation of the unit cell parameters were performed using Jana2006 programs [38] with the Le Bail method [39]. Reliable structural parameters are demonstrated based on $R_p$ = 16.8%, $R_{wp}$ = 18.3%.

**Resistance measurements.**

The low-temperature environment was provided by a helium-4 closed-cycle refrigerator. The electrical resistivity was measured using a van der Pauw lock-in technique at 17 Hz with Lakeshore cryotronics 370 AC resistance bridge. A current of 316 μA was used to measure the resistivity at low temperatures. The temperature was recorded by a resistance thermometer. The ramping speed was at 0.1 K/min during the cooling and warming procedures controlled by Lakeshore Model 350 Cryogenic temperature controller.

**Critical current measurements.**

Keithley 6221 and Keithley 2182A were employed for the critical current measurements. Measurements of *I–V* characteristics were performed in a pulsed mode to minimize Joule heating, where the duration of the pulsed current was 10 (1) ms and the repetition rate was 2 s [40]. Max current was limited to 80 mA to prevent any possible damage to the electrodes.

**Quasiparticle scattering spectroscopy measurements.**

Details of quasiparticle scattering spectroscopy measurements under pressure can be found elsewhere [33, 41]. With careful treatment of the Pt slices, the quasiparticle scattering radius at the Pt/CeH$_9$ interface in the DAC can be reduced to below 5 $\mu$m, which is smaller than the extremity of the tip in the hard quasiparticle scattering spectroscopy. The voltage responses to the DC current were measured at three small current steps ($\Delta I$ is less than 0.5% of the whole curve), and the slope was calculated to obtain the differential conductance value. The corresponding bias voltage in the junction was estimated as an average voltage of the three values. All values of differential conductance were obtained from the 3-point moving average to minimize the effects of the electromotive force in the circuit. The observed nonlinear background was analyzed by the polynomial fitting in the voltage range that is higher than that where the zero-bias conductance peak is observed.

**Raman scattering measurements.**

The Raman spectra were measured using a 320 Princeton Instrument IsoPlane spectrograph equipped with a liquid nitrogen-cooled CCD PyLoN detector. Optigrate's BragGrate notch filters and bandpass filters were used, making Raman spectroscopy measurements close to the Rayleigh line possible for both stokes and anti-stokes scattering down to ± 5 cm$^{-1}$. The 532 and 660 nm excitation wavelengths illuminate a spot with a diameter of approximately ~2 μm. The Raman spectra were measured with a typical laser power of 0.65-1 mW to reduce laser heating effects.

**Data availability**

All data that support the findings of this study are available from the corresponding author upon reasonable request.

**Acknowledgments**

Z.Y.C. acknowledges the Special Construction Project Fund for Shandong Province Taishan Scholars. The work at HPSTAR was supported by the National Key R&D Program of China (Grant no. 2018YFA0305900). L.C.C. and X.J.C. acknowledge the support from the Basic Research Program of Shenzhen (Grant No. JCYJ20200109112810241) and the Shenzhen Science and Technology Program (Grant No. KQTD20200820113045081). We appreciate APCTP and BrainLink for their hospitality during the completion of this work. The work at Sungkyunkwan University was supported by the National Research Foundation (NRF) of Korea through a grant funded by the Korean Ministry of Science and ICT (MIST) (No. 2021R1A2C2010925, No. RS-2023-00220471). Additionally, this research was supported by the Sungkyunkwan University and the BK21 FOUR (Graduate School Innovation) program, and Basic Science Research Program (No.2022R1I1A1A01071050) funded by the Ministry of Education (MOE, Korea) and National Research Foundation of Korea (NRF).

**Author contributions**

T.P. and X.J.C. conceived the project and led this study from Korea and China, cooperatively. X.J.C. was responsible for the overall direction, planning and integration among different research units in HPSTAR, where the cell preparation, sample synthesis, resistance, x-ray diffraction, and Raman scattering measurements were conducted. Under the supervision of X.J.C., Z.Y.C. prepared the cell and detected the superconducting transition, F.A.G., P.D., Z.Y.C., J.F.Y., J.K. and Y.J.L. performed the laser heating and Raman spectroscopy measurements, Z.Y.C., L.C.C., and G.H. performed x-ray diffraction studies, and L.C.C. analyzed the structural data. The superconducting critical current and quasiparticle scattering spectroscopy measurements were carried out at Sungkyunkwan University by S.M.C., Z.Y.C., H.J., and S.-G.J under the supervision of T.P. The experimental data analysis was performed by Z.Y.C., S.M.C., H.J., L.C.C., L.Y., S.-G.J., and T.P. All authors contributed to the discussion of the results. Z.Y.C. drafted the manuscript and X.J.C. and T.P. edited the manuscript with the key inputs from P.D. and H.J.

**Competing interests**

The authors declare no competing interests. .